\documentstyle[proceedings,numreferences,epsf]{crckapb}

\newcommand{\Mpc}{\hbox{\,h$^{-1}$\,Mpc}}

\begin{opening}
\title{REDSHIFT SURVEYS WITH 2DF}
\author{MATTHEW COLLESS$^\ast$}
\institute{Mount Stromlo \& Siding Spring Observatories, Australia}
\author{BRIAN BOYLE$^\dagger$}
\institute{Anglo-Australian Observatory, Australia}
\end{opening}

\begin{document}

\footnotetext[1]{ 
{\sl On behalf of the 2dF Galaxy Redshift Survey Team:} Matthew
Colless (MSSSO), Richard Ellis (IoA), Joss Bland-Hawthorn (AAO),
Russell Cannon (AAO), Shaun Cole (Durham), Chris Collins (LJMU),
Warrick Couch (UNSW), Gavin Dalton (Oxford), Simon Driver (UNSW),
George Efstathiou (IoA), Simon Folkes (IoA), Carlos Frenk (Durham),
Karl Glazebrook (AAO), Nick Kaiser (IfA), Ofer Lahav (IoA), Ian Lewis
(AAO), Stuart Lumsden (AAO), Steve Maddox (IoA), John Peacock (ROE),
Bruce Peterson (MSSSO), Will Sutherland (Oxford), Keith Taylor (AAO)}
\footnotetext[2]{
{\sl On behalf of the 2dF QSO Redshift Survey Team:} Brian Boyle
(AAO), Scott Croom (Durham), Lance Miller (Oxford), Mike Read (ROE),
Tom Shanks (Durham), Robert Smith (IoA)}

\section{Introduction}

This IAU Joint Discussion proposes to address the subject of redshift
surveys in the 21st century. This paper, however, deals with two major
new redshift surveys that those involved sincerely hope will be
completed in the 20th century. Nonetheless, these surveys are relevant
to the topic of the meeting, as they clearly foreshadow the scope and
style of redshift surveys, if not in the coming millennium, at least in
the coming decade.

The surveys are being carried out with the new Two Degree Field (2dF)
facility on the Anglo-Australian Telescope (AAT), a 400-fibre
multi-object spectrograph with the capability, as described in
Section~2, to increase the size of redshift surveys by an order of
magnitude over current best efforts. The main scientific goals, survey
strategy and some preliminary results from the 2dF Galaxy Redshift
Survey are outlined in Section~3, while Section~4 similarly describes
the 2dF QSO Redshift Survey. Further information can be found on the
WWW at {\tt http://www.aao.gov.au/2df/} for the 2dF facility, at {\tt
http://msowww.anu.edu.au/$\sim$colless/2dF/} for the galaxy survey and
at {\tt http://www.aao.gov.au/local/www/rs/qso\_surv.html} for the QSO
survey.

\section{The 2dF Facility}

\begin{figure}
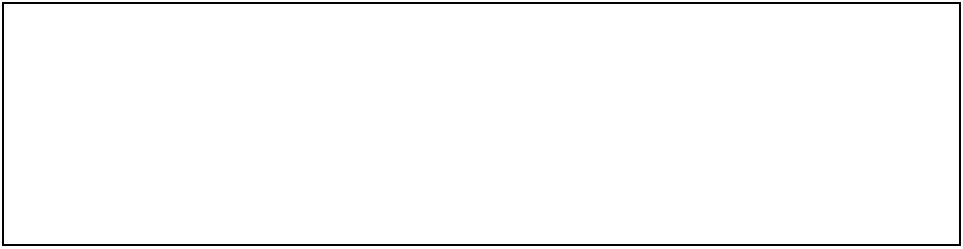

\centering 
\fbox{\parbox{\textwidth}{\hfill\vspace*{3cm}}}
\caption{Schematic of the 2dF facility.
\label{fig:2dFtopend}}
\end{figure}

The 2dF facility on the AAT consists of several major components,
shown schematically in Figure~\ref{fig:2dFtopend}: a new prime focus
optical assembly, comprising an atmospheric dispersion compensator and
field corrector; a pair of field plates, each with 400 positionable
fibres, which can be tumbled between the focus location and a
configuration location; a robotic fibre positioner which can rapidly
re-configure the 400 fibres; two spectrographs, each dealing with the
light from 200 fibres; and a top-end ring assembly to carry these
various components.

\begin{figure}
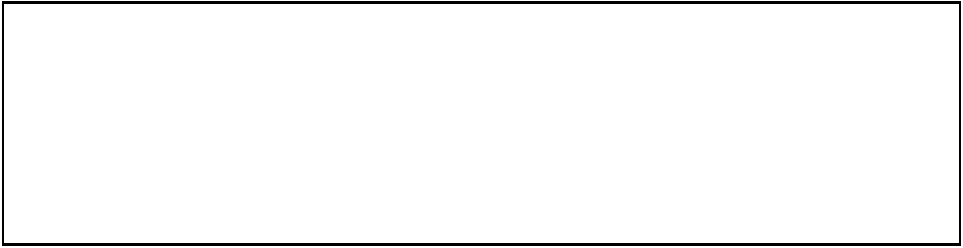

\centering 
\fbox{\parbox{\textwidth}{\hfill\vspace*{3cm}}}
\caption{One of the two fieldplates, showing 400 fibres positioned in
the focal plane.
\label{fig:plate400}}
\end{figure}

2dF obtained its first scientific data in November 1996. For the next
nine months it operated with only 200 fibres and one spectrograph
while the system was tested and commissioned and the second
spectrograph completed. In September 1997 the full 400 fibres were
installed (see Figure~\ref{fig:plate400}) and for the first time 400
spectra were obtained simultaneously using the two-spectrograph
system.

\section{The 2dF Galaxy Redshift Survey}

\subsection{Goals of the Survey}

\begin{figure}
\centering 
\parbox{\textwidth}{\epsfxsize=\textwidth \epsfbox{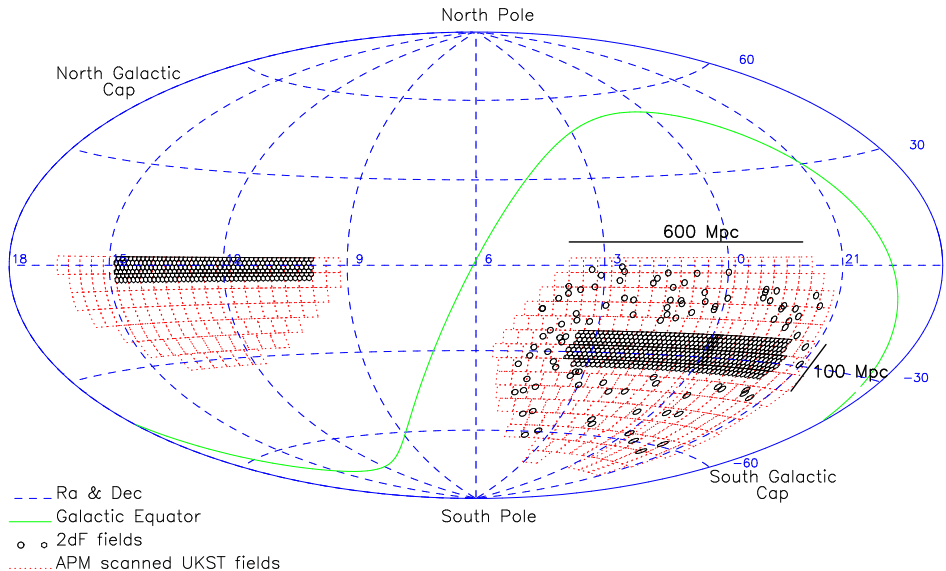}}
\caption{The 2dF survey fields (small circles) superimposed on the APM
survey area (dotted outlines of Sky Survey plates). There are
approximately 140,000 galaxies in the 75$^\circ$$\times$12.5$^\circ$
South Galactic Hemisphere (SGH) strip centred on the South Galactic
Pole, 70,000 galaxies in the 65$^\circ$$\times$7.5$^\circ$ North
Galactic Hemisphere (NGH) equatorial strip, and 40,000 galaxies in the
100 random 2dF fields scattered over the entire southern region of the
APM galaxy survey. 
\label{fig:skyplot}}
\end{figure}

The observational goal of the 2dF Galaxy Redshift Survey is to secure
high quality spectra and redshifts for 250,000 galaxies brighter than
b$_J$=19.5 (extinction-corrected), with a deeper extension to R=21
making best use of good conditions. The brighter galaxies cover an
area of 1700 square degrees selected from both the southern galactic
hemisphere APM galaxy survey and an extension into the north galactic
hemisphere equatorial region. The arrangement of 2dF survey fields is
shown in Figure~\ref{fig:skyplot}.

We aim to use the survey to address a variety of fundamental problems
in galaxy formation and cosmology. The main scientific goals include:

1. Accurate measurements of the power spectrum of galaxy clustering on
scales of up to a few hundred Mpc, allowing direct comparisons with
microwave background anisotropy measurements of fluctuations on the
same spatial scales.

2. Measurements of the distortion of the clustering pattern in
redshift space, providing constraints on the cosmological density
parameter $\Omega$ and the spatial distribution of dark matter.

3. A determination of variations in the spatial and velocity
distributions of galaxies as a function of luminosity, type and
star-formation history, providing important constraints on models of
galaxy formation.

4. Measurements of the evolution of the galaxy luminosity function,
clustering amplitude and star formation rates out to a redshift of
$z$$\sim$0.5. This will be accomplished by combining the bright
b$_J$$<$19.5 survey with the faint R$<$21 survey, which is comparable
in size to the largest current local surveys.

5. Investigations of the morphology of galaxy clustering and the
statistical properties of the fluctuations---for example, whether the
initial fluctuations are Gaussian as predicted by inflationary models
of the early universe.

6. A study of clusters and groups of galaxies in the redshift survey,
in particular the measurement of infall in clusters and dynamical
estimates of cluster masses at large radii.

7. Application of novel techniques to classify the uniform sample of
250,000 spectra obtained in the survey, thereby obtaining a
comprehensive inventory of galaxy types as a function of spatial
position within the survey.

\subsection{Survey Status}

As of mid-1997 we have taken nearly 2000 redshifts for the survey.
These were obtained in commissioning time on the instrument using only
one of the two spectrographs and 200 fibres. Later this year we will
begin observations with all 400 fibres and the survey will enter
production-line operation. We expect to complete the survey
observations by the end of 1999.

\begin{figure}
\centering 
\parbox{\textwidth}{\epsfxsize=0.7\textwidth \epsfbox{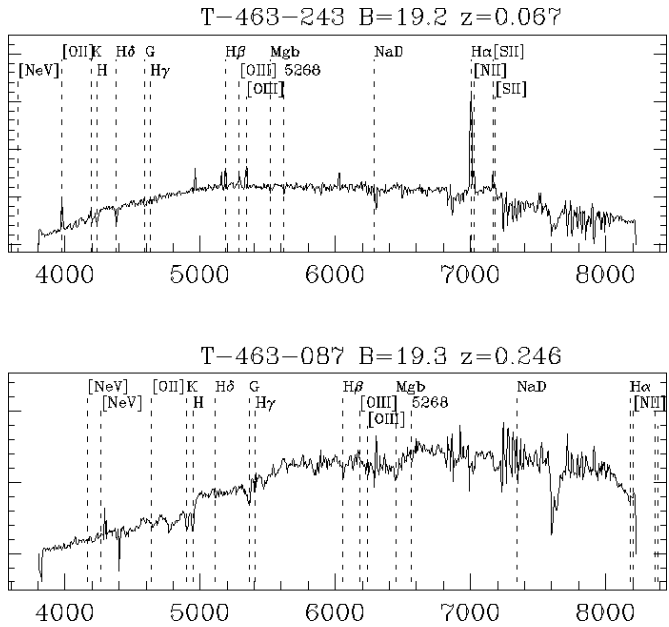}}
\caption{Example spectra from the survey: a b$_J$=19.2 emission-line
galaxy at $z$=0.067 and a b$_J$=19.3 absorption-line galaxy at
$z$=0.246.
\label{fig:specs}}
\end{figure}

Two example spectra taken in the commissioning period are shown in
Figure~\ref{fig:specs}. One is a b$_J$=19.2 emission-line galaxy at
$z$=0.067 and the other is a b$_J$=19.3 absorption-line galaxy at
$z$=0.246. The quality target for the survey spectra is a S/N of at
least 10 per 2-pixel resolution element (FWHM$\approx$9\AA); most
spectra will easily exceed this target. 

\begin{figure}
\centering 
\parbox{\textwidth}{\epsfxsize=0.6\textwidth \epsfbox{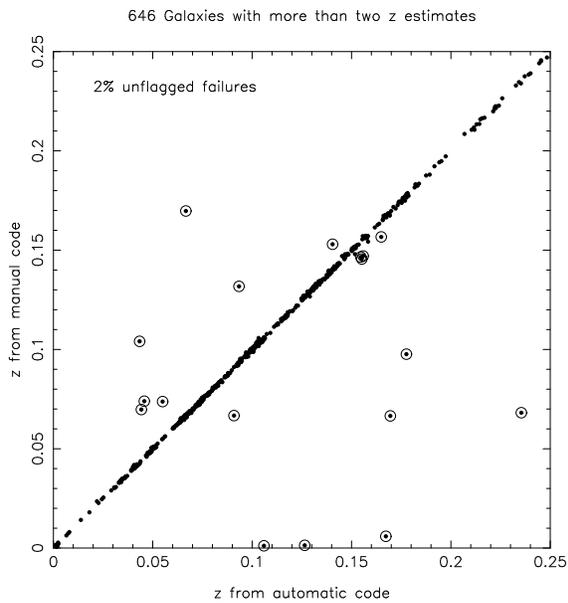}}
\caption{A comparison of redshifts determined automatically and via
visual inspection of the spectra.
\label{fig:z1-z2}}
\end{figure}

This minimum S/N permits reliable automatic spectral classification
and redshift measurement. Employing both a standard cross-correlation
and line-fitting code and a new code which uses Principal Component
Analysis and $\chi^2$-fitting to simultaneously classify the spectrum
and measure its redshift (Glazebrook et~al.\ 1997), we find we achieve
a very high level of reliability. A comparison of the redshifts
obtained from these codes with redshifts determined semi-manually via
visual inspection (Figure~\ref{fig:z1-z2}) shows a very low level of
failures in the automatic algorithms, even with the variable-quality
commissioning data.

\subsection{Preliminary Results}

\begin{figure}
\centering 
\parbox{\textwidth}{\epsfxsize=\textwidth \epsfbox{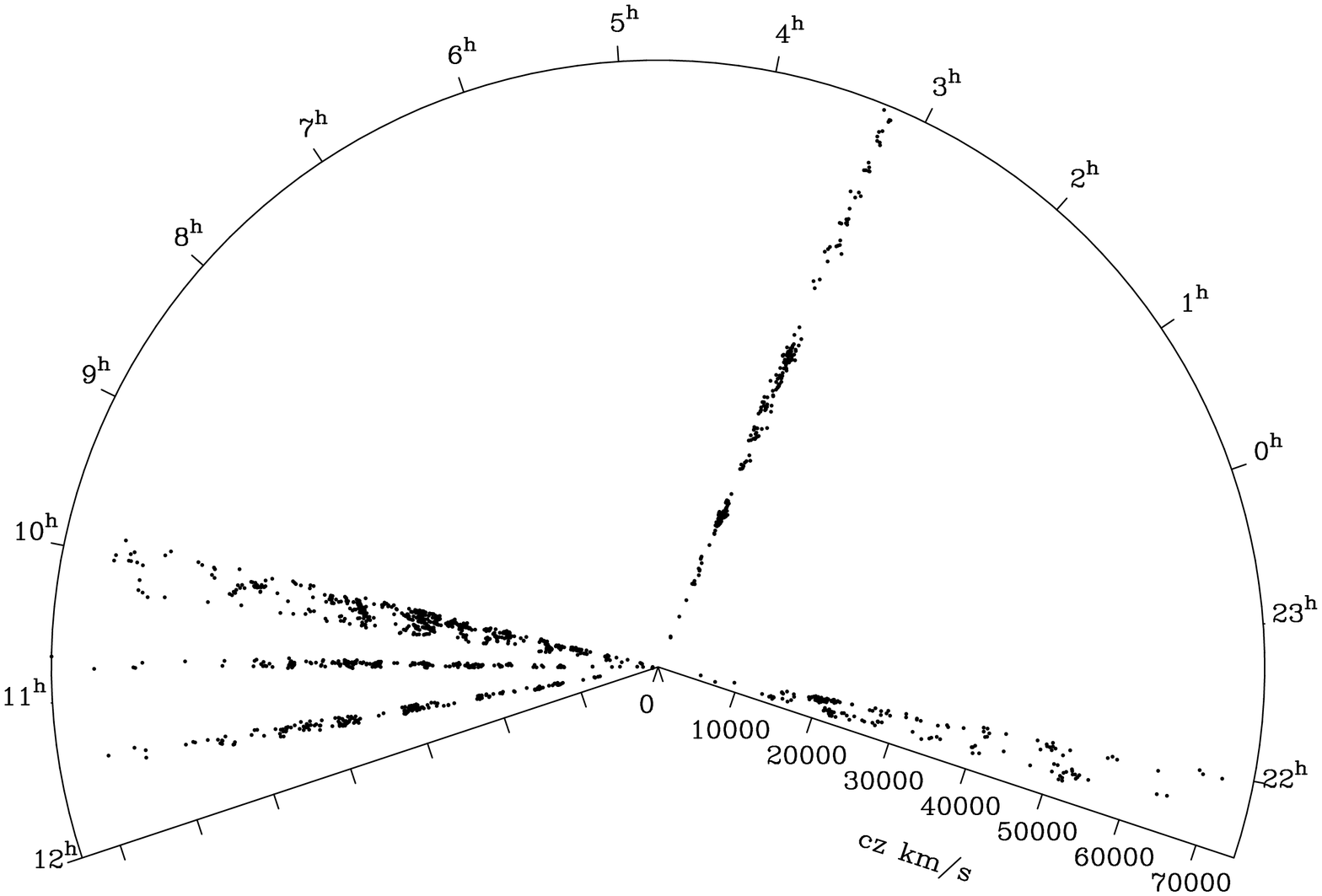}}
\vspace*{12pt}
\parbox{\textwidth}{\epsfxsize=\textwidth \epsfbox{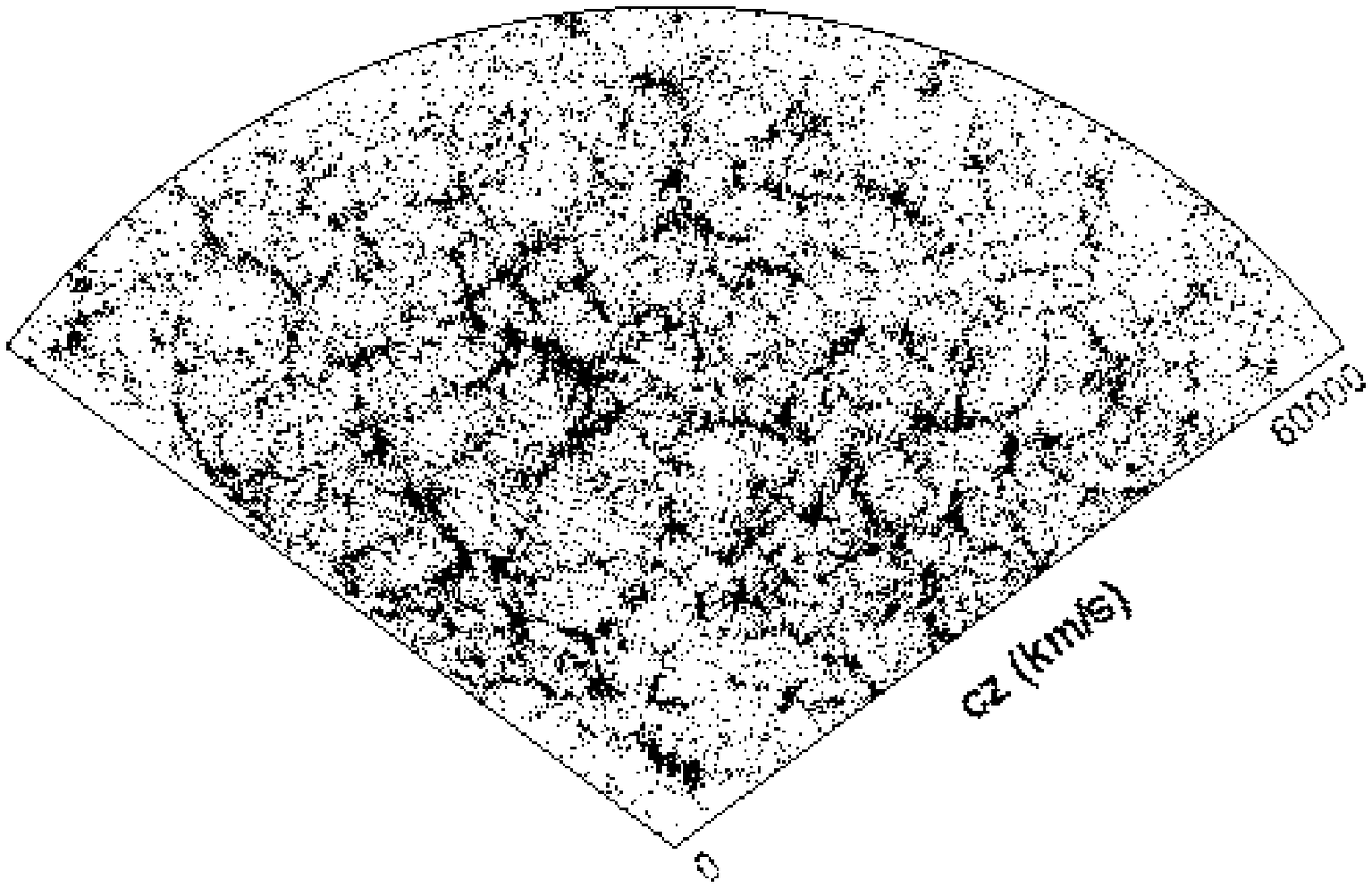}}
\caption{(a)~A cone plot of the survey redshifts to date, combining
NGH and SGH strips and including $\sim$1500 galaxies (less than 1\% of
the full sample). (b)~A simulated cone plot of a single 2$^\circ$-wide
declination strip comprising 34,000 galaxies (less than 15\% of the
full sample).
\label{fig:zslice}}
\end{figure}

Figure~\ref{fig:zslice}a is a cone plot, combining NGH and SGH strips,
showing the galaxies with redshifts measured to date---this is less
than 1\% of the full sample. For comparison, Figure~\ref{fig:zslice}b
shows a simulated cone plot of a single 2$^\circ$ declination slice in
the redshift survey; even this slice, which contains 34,000 galaxies,
represents less than 15\% of the survey.

\begin{figure}
\centering 
\parbox{\textwidth}{\epsfxsize=0.8\textwidth \epsfbox{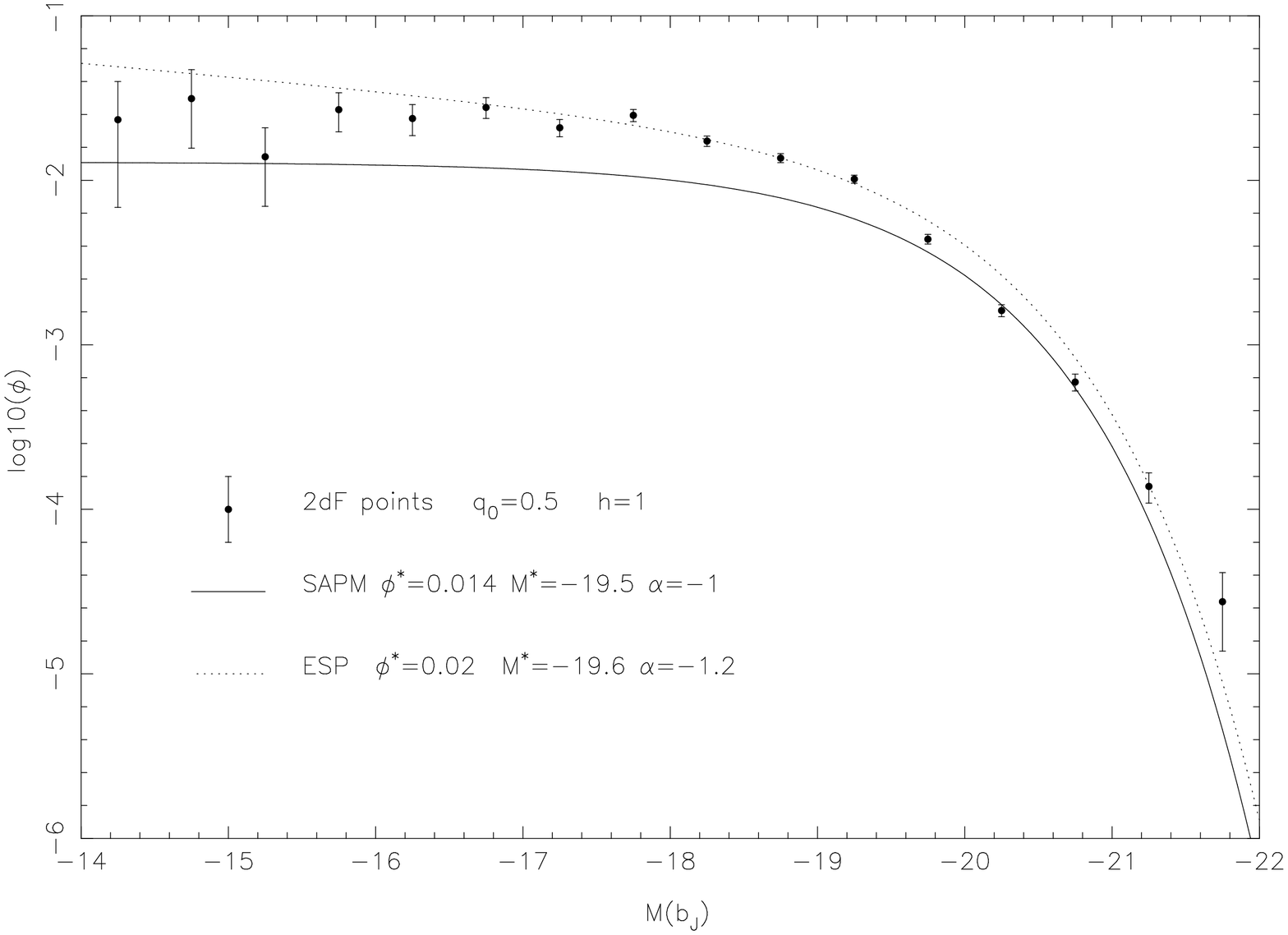}}
\caption{A preliminary galaxy luminosity function from the 2dF survey,
compared to previous determinations by the APM-Stromlo and ESO Slice
Project surveys.
\label{fig:phi}}
\end{figure}

Although so far we have observed too few fields to effectively address
questions of large-scale structure, we do have a sufficiently large
sample of redshifts to begin to look at the properties of local
galaxies. Figure~\ref{fig:phi} shows a preliminary determination of
the galaxy luminosity function (LF) at a mean redshift of
$\bar{z}$=0.1. This LF is in good agreement with the LFs from the
Autofib survey (Ellis et~al.\ 1997) and the ESO Slice Project (Zucca
et~al.\ 1997), and gives an overall galaxy density about 50\% higher
than that obtained in the Stromlo-APM survey (Loveday et~al.\ 1992).
It is important to note that we are reaching more than 5 magnitudes
below $L^*$ even with this small sample, and that we should be able to
tightly constrain the faint end of the local luminosity function for
each morphological or spectral type with the full sample.

\section{The 2dF QSO Redshift Survey}

\subsection{Goals of the Survey}

The observational goal of the QSO redshift survey is to obtain spectra
for $\sim$30000 $B$$<$21 QSOs in two declination strips at the South
Galactic Pole and in an equatorial region at the North Galactic cap.
The QSO survey will probe the largest scales in the universe
(10\Mpc~$<r<$~1000\Mpc) over a wide range in redshift space
($0.3<z<2.9$), and is thus complementary to the galaxy survey. The
primary scientific aims of the survey are:

1. To obtain the primordial fluctuation power spectrum out to COBE
scales.

2. To determine the rate of QSO clustering evolution in the non-linear
and linear regimes, and hence obtain new limits on the value of
$\Omega$ and $b$ (Croom \& Shanks 1996).

3. To apply geometric methods to measure $\Lambda$ (Ballinger et~al.\
1996).

4. To identify large statistical samples of unusual classes of QSOs
(e.g.\ BALs) or absorption line systems (e.g.\ damped Ly$\alpha$
systems).

Observations that radio-quiet QSOs exist in average galaxy cluster
environments ($r_0$$\sim$5\Mpc) at low-to-moderate redshifts (Smith
et~al.\ 1995 and references therein) demonstrate that radio-quiet QSOs
can be used to provide important information on the structure of the
Universe at the largest scales.

\subsection{Survey Strategy and Status}

\begin{figure}
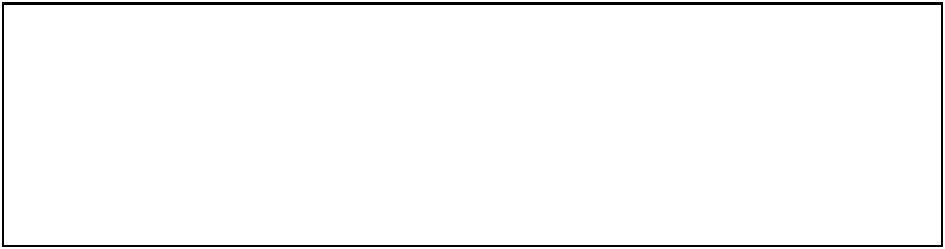

\centering 
\framebox[\textwidth]{\parbox{\textwidth}{\hfill\vspace*{3cm}}}
\caption{The UVX-selected objects for the QSO 2dF input catalogue. The
RA and Dec of all UVX-selected objects are shown in the two
$75^{\circ} \times 5^{\circ}$ strips of the catalogue. Increasing
contamination by Galactic sub-dwarfs at lower Galactic latitudes gives
an obvious gradient ($<$50\%) in number density along the survey
strips. Completeness estimates are based on our success at recovering
previously identified QSOs.
\label{fig:uvxcat}}
\end{figure}

The 2dF QSO survey covers an area of 740~deg$^2$ comprising two
$75^{\circ} \times 5^{\circ}$ strips on the sky (small areas
surrounding bright stars have been excluded from the analysis). QSO
candidates were selected from APM measurements of UK Schmidt $U$, $J$
and $R$ plates/films, on the basis of their anomalous position in the
stellar $U$$-$$B$/$B$$-$$R$ colour--colour diagram. The vast majority
($>$85\%) of the candidates are selected solely on the basis of their
$U$$-$$B$ colour (sensitive to QSOs with $z$$<$2.2), but the
additional use of the $R$ magnitude minimises contamination from
Galactic stars and extends the redshift range over which QSOs can be
selected to $z$$\sim$2.9. The initial UVX-selected catalogue is shown
in Figure~\ref{fig:uvxcat}. In total over 150 $U$, $J$ and $R$ plates
(comprising 30 UKST fields) were used to compile the input catalogue.
Great care was taken to keep photometric variations at $<$0.1~mag
level over the entire catalogue (Smith 1997).

The QSO 2dF redshift survey fields form part of the area of sky
surveyed by the galaxy 2dF survey, and so the objects in the QSO and
galaxy 2dF redshift surveys are being observed simultaneously with
2dF. This leads to significant gains in both efficiency (the combined
survey can be carried out in 20 nights less than if the surveys had
been separate) and completeness (the QSO survey will pick up compact
blue galaxies, the galaxy survey will identify extended low-redshift
QSOs).

\begin{figure}
\centering 
\parbox{\textwidth}{\epsfxsize=\textwidth \epsfbox{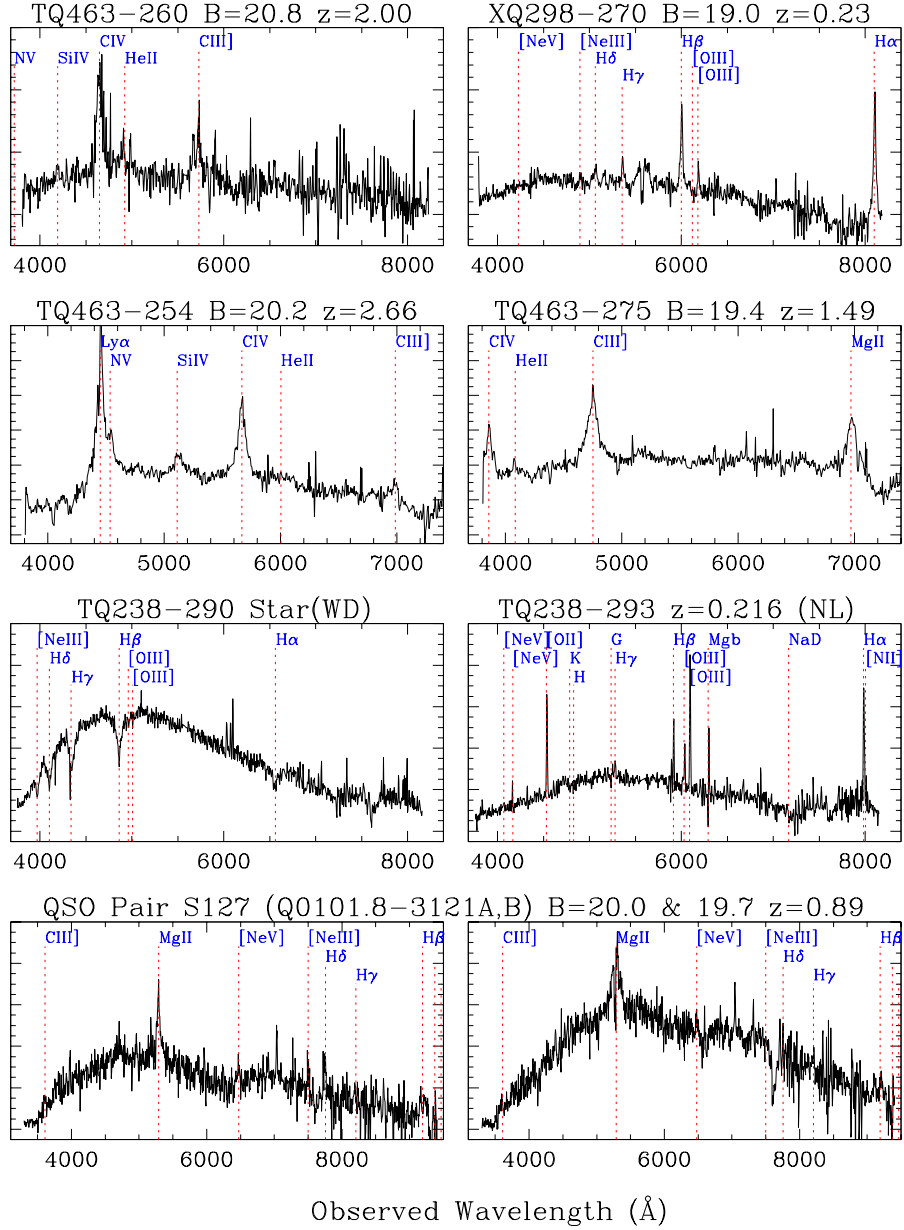}}
\caption{Representative spectra from the 2dF QSO redshift survey. The
first six spectra were obtained with the 2dF and comprise 4 QSOs, a
white dwarf and a narrow-emission-line galaxy. The latter two classes
of object are some of the typical `contaminants' of the UVX-selected
sample. The bottom pair of objects are a close pair of QSOs (20~arcsec
separation) observed with the MSSSO 2.3m.
\label{fig:qspec}}
\end{figure}

To date (October 1997), almost 800 QSO candidates in the 2dF QSO
survey have been observed, comprising over 400 QSOs. Most of these
have been observed with 2dF, but over 100 bright ($B$$<$18) QSOs have
been identified with the UKST FLAIR system and 20 QSOs in close pairs
(separations less than 20~arcsec) have been discovered with the MSSSO
2.3m telescope. Representative spectra from the QSO survey are shown
in Figure~\ref{fig:qspec}.

We hope to complete the QSO 2dF redshift survey over the next two
years, providing an invaluable resource with which to study the
large-scale structure of the Universe out to $z$$\sim$3. Progress of
the survey can be monitored through the WWW site (see Introduction),
and by-products of the survey (CCD calibration, input catalogue) will
be available at regular intervals.


\begin{thebibliography}{99}
\bibitem{}
Ballinger W.E., Peacock J.A., Heavens A.F., 1996, MNRAS, 282, 877
\bibitem{}
Colless M.M., 1996, Wide Field Spectroscopy and the Universe, {\it
Wide Field Spectroscopy}, eds Kontizas M., Kontizas E., Kluwer,
pp.227--240
\bibitem{}
Croom S.M., Shanks T., 1996, MNRAS, 281, 893
\bibitem{}
Ellis R.S., Colless M.M., Broadhurst T.J., Heyl J.S., Glazebrook K.,
1996, MNRAS, 280, 235
\bibitem{}
Glazebrook K., Offer A.R., Deeley K., 1997, MNRAS, in press
\bibitem{}
Loveday J., Peterson B.A., Efstathiou G., Maddox S.J., 1992, ApJ, 390,
338
\bibitem{}
Smith R.J., 1997, PhD Thesis, University of Cambridge
\bibitem{}
Smith R.J., Boyle B.J., Maddox S.M., 1995, MNRAS, 277, 270
\bibitem{}
Zucca E., Zamorani G., Vettolani G., Cappi A., et~al., 1997, A\&A, in
press 
\end{thebibliography}
\end{document}